\documentclass[prl,aps]{revtex4}
\usepackage{graphicx}
\usepackage{multirow}
\usepackage{amssymb}
\def\lsim{\mathrel{\raise.3ex\hbox{$<$\kern-.75em\lower1ex\hbox{$\sim$}}}}
\def\gsim{\mathrel{\raise.3ex\hbox{$>$\kern-.75em\lower1ex\hbox{$\sim$}}}}
\newcommand{\ed}{\end{document}}
\DeclareMathAlphabet{\mathsc}{OT1}{cmr}{m}{sc}

\newcommand{\nsi}{Bergmann:1998ft,%
Ota:2001pw,%
Ota:2002na,%
Honda:2006gv,%
Kitazawa:2006iq,%
Friedland:2006pi,%
Blennow:2007pu,%
GonzalezGarcia:2001mp,%
Huber:2001zw,%
Gago:2001xg,%
Huber:2002bi,%
Campanelli:2002cc,%
Bueno:2000jy,%
Kopp:2007mi,%
Adhikari:2006uj,%
Ribeiro:2007ud,%
Kopp:2007ne}

\begin{document}
\preprint{IFIC/08-17}

\title{Can OPERA help in constraining neutrino non-standard interactions?}

\author{A. Esteban-Pretel}
\affiliation{ AHEP Group, Institut de F\'{\i}sica Corpuscular --
  C.S.I.C./Universitat de Val{\`e}ncia \\
  Edificio Institutos de Paterna, Apt 22085, E--46071 Valencia, Spain}

\author{P. Huber}
\affiliation{Theory Division, Department of Physics, CERN\\
CH-1211 Geneva 23, Switzerland}

\affiliation{Institute for Particle, Nuclear and Astronomical
  Sciences,\\
Physics Department, Virgina Tech, Blacksburg, VA 24062, USA}

\author{J.W.F. Valle}
\affiliation{ AHEP Group, Institut de F\'{\i}sica Corpuscular --
  C.S.I.C./Universitat de Val{\`e}ncia \\
  Edificio Institutos de Paterna, Apt 22085, E--46071 Valencia, Spain}

\date{\today}

\begin{abstract}

  We study how much the unique ability of the OPERA experiment to
  directly detect $\nu_\tau$ can help in probing new, non-standard
  contact interactions of the third family of neutrinos. We perform a
  combined analysis of future, high-statistics MINOS and OPERA
  data. For the case of non-standard interactions in $\nu_\mu$ to
  $\nu_e$ transitions we also include the impact of possible
  DoubleCHOOZ data. In all cases we find that the $\nu_\tau$ sample of
  OPERA is too small to be statistically significant, even if one
  doubles the nominal exposure of OPERA to $9\times10^{19}$
  pot. OPERA's real benefit for this measurement lies in its very high
  neutrino energy and hence very different $L/E$ compared to MINOS.

\end{abstract}

\maketitle
\section{Introduction}

The confirmation of the neutrino oscillation interpretation of solar
and atmospheric neutrino data by reactor~\cite{araki:2004mb} and
accelerator~\cite{Ahn:2006zz,Michael:2006rx} neutrino experiments
brings a unique picture of neutrino physics in terms of three-neutrino
oscillations~\cite{Maltoni:2004ei}, leaving little room for other
non-standard neutrino properties~\cite{pakvasa:2003zv}.
Nevertheless, it has long been recognized that any gauge theory of
neutrino mass generation inevitably brings in dimension-6 non-standard
neutrino interaction (NSI) terms.  Such sub-weak strength operators
arise in the broad class of seesaw-type models, due to the non-trivial
structure of charged and neutral current weak
interactions~\cite{schechter:1980gr}. Similarly, NSI also appear in
radiative models of neutrino mass.
They can be of two types: flavor-changing (FC) and non-universal (NU)
and their strength $\varepsilon G_F$ is highly model-dependent but may
lie within the sensitivities of currently planned experiments.
The presence of NSI leads to possibly new resonant effects in the
propagation of astrophysical
neutrinos~\cite{Valle:1987gv,nunokawa:1996tg,nunokawa:1996ve,grasso:1998tt,EstebanPretel:2007yu}
and it is interesting to scrutinize their possible role in the
propagation of laboratory neutrinos. With neutrino oscillation physics
entering the precision age~\cite{Bandyopadhyay:2007kx,Nunokawa:2007qh}
it becomes an important challenge to investigate the role of NSI in
future  terrestrial neutrino oscillation experiments.
 
The interplay of oscillation and neutrino non-standard interactions
(NSI) was studied in~\cite{Grossman:1995wx} and subsequently it was
shown~\cite{Huber:2001de,Huber:2002bi} that in the presence of NSI it
is very difficult to disentangle genuine oscillation effects from
those coming from NSI. The latter may affect production, propagation
and detection of neutrinos and in general these three effects need not
be correlated.  It has been shown that in this case cancellations can
occur which make it impossible to separate oscillation from NSI
effects.  Subsequently it was discovered that the ability to detect
$\nu_\tau$ may be crucial in order to overcome that
problem~\cite{Campanelli:2002cc}, though this method requires
sufficiently large beam energies to be applicable.  Barring the
occurrence of fine-tuned cancellations, NSI and oscillations have very
different $L/E$ dependence. Therefore, combining different $L/E$ can
be very effective in probing the presence of NSI. The issue of NSI and
oscillation in neutrino experiments with terrestrial sources has been
studied in a large number of publications~\cite{\nsi}.
In~\cite{Blennow:2007pu} it was shown that MINOS~\cite{Michael:2006rx}
on its own is not able to put new constraints on NSI parameters. On
the other hand, in~\cite{Friedland:2006pi} the combination of
atmospheric data with MINOS was proven to be effective in probing at
least some of the NSI parameters. Since matter effects are relatively
small in MINOS, its main role in that combination is to constrain the
vacuum mixing parameters.

The question we would like to address here is whether the combination
of MINOS and OPERA~\cite{Guler:2000bd} can provide useful information
on NSI. OPERA has recently seen the first events in the emulsion cloud
chamber~\cite{OPERAfirst} and hence it appears timely to ask this
question. The idea is that OPERA will be able to detect $\nu_\tau$ and
has a very different $L/E$ than MINOS.  Both factors are known to help
distinguishing NSI from oscillation effects.  Clearly, much larger
improvements on existing sensitivities are expected from superbeam
experiments like T2K~\cite{Itow:2001ee} and
NO$\nu$A~\cite{Ayres:2004js} especially in combination with reactor
neutrino experiments like
DoubleCHOOZ~\cite{Ardellier:2004ui,Ardellier:2006mn} or Daya
Bay~\cite{Guo:2007ug}, see Ref.~\cite{Kopp:2007ne}.  In this letter we
will focus on the simple case where NSI only affects neutrino
propagation.

\section{Basic Setup}

Adding NSI into the propagation of neutrinos yields the following
evolution Hamiltonian
\begin{equation}
\label{eq:hamiltonian}
\mathcal{H}=\frac{1}{2E}U\left(\begin{array}{ccc}
0&0&0\\
0&\Delta m^2_{21}&0\\
0&0&\Delta m^2_{31}
\end{array}\right)U^\dagger +
\frac{1}{2E}\left(\begin{array}{ccc}
V&0&0\\
0&0&0\\
0&0&0
\end{array}\right)+ \frac{V}{2E}\left(\begin{array}{ccc}
0&0&\varepsilon_{e\tau}\\
0&0&0\\
\varepsilon_{e\tau}&0&\varepsilon_{\tau\tau}
\end{array}\right) \,,
\end{equation}
where we have made use of the fact that all $\varepsilon_{x\mu}$ are
fairly well constrained and hence are expected not to play a
significant role at leading order. The effect of $\varepsilon_{ee}$ is a
re-scaling of the matter density and all experiments considered here
are not expected to be sensitive to matter effects. Hence we will set
$\varepsilon_{ee}= 0$. Note, that the $\varepsilon$ as defined here, are
effective parameters. At the level of the underlying Lagrangian
describing the NSI, the NSI coupling of the neutrino can be either to
electrons, up or down quarks. From a phenomenological point of view,
however, only the (incoherent) sum of all these contributions is
relevant. For simplicity, we chose to normalize our NSI to the
electron abundance.  This introduces a relative factor of 3 compared
to the case where one normalizes either to the up or down quark
abundance (assuming an isoscalar composition of the Earth), {\it i.e.}
the NSI coupling to only up or down quark would need to be 3 times as
strong to produce the same effect in oscillations. Since both
conventions can be found in the literature, care is required in making
quantitative comparisons.

There are two potential benefits beyond adding statistics from
combining the data from MINOS and OPERA: First, OPERA can detect
$\nu_\tau$ which, in principle, allows to directly access any effect
from $\varepsilon_{x\tau}$. Moreover, although the baseline is the same,
the beam energies are very different $\langle E \rangle \simeq 3
\,\mathrm{GeV}$ for MINOS, whereas $\langle E \rangle \simeq 17
\,\mathrm{GeV}$ for OPERA.

\subsection{Experiments}

All numerical simulations have been done using the GLoBES
software~\cite{Huber:2004ka,Huber:2007ji}. In order to include the
effects of the NSI we have customized the package by adding a new
piece to the Hamiltonian as shown in equation~\ref{eq:hamiltonian}. We
have considered three different experiments: MINOS, OPERA and
DoubleCHOOZ, the main characteristics of which are summarized in
table~\ref{tab:exp}.

\begin{table}[!b]
  \begin{tabular}{l@{\hspace{0.5cm}}rcl@{\hspace{0.5cm}}r@{\hspace{0.5cm}}c@{\hspace{0.5cm}}c@{\hspace{0.5cm}}rcl}
    \hline
    \hline
    Label & & $L$ & & $\langle E_\nu \rangle$ & power & $t_{run}$ & \multicolumn{3}{c}{channel} \\
    \hline
    \hline
    MINOS$_2$ (M2) & 735 & km & & 3 GeV & $5\times 10^{20}$ pot/yr & 5 yr & $\nu_\mu$ & $\to$ & $\nu_{e,\mu}$ \\
    \hline
    OPERA (O) & 732 & km & & 17 GeV & $4.5\times 10^{19}$ pot/yr & 5 yr & $\nu_\mu$ & $\to$ & $\nu_{e,\mu,\tau}$ \\
    \hline
    \multirow{2}{*}{DoubleCHOOZ (DC)} & 0.2 & km & (near) & \multirow{2}{*}{4 MeV} & \multirow{2}{*}{$8.4$ GW} & \multirow{2}{*}{5 yr} & \multirow{2}{*}{$\bar\nu_e$} & \multirow{2}{*}{$\to$} & \multirow{2}{*}{$\bar\nu_e$} \\
    & 1.05 & km & (far) & & & & & \\
    \hline
    \hline
  \end{tabular}
  \caption{\label{tab:exp}
    Main parameters of the experiments under study.}
\end{table}

MINOS is a long baseline neutrino oscillation experiment using the
NuMI neutrino beam, at FNAL. It uses two magnetized iron calorimeters.
One serves as near detector and is located at about $1\,\mathrm{km}$
from the target, whereas the second, larger one is located at the
Soudan Underground Laboratory at a distance of $735\,\mathrm{km}$ from
the source. The near detector is used to measure the neutrino beam
spectrum and composition. The near/far comparison also mitigates the
effect of cross section uncertainties and various systematical errors.
In our simulations, based
on~\cite{Huber:2004ug,Ables:1995wq,NUMIL714}, we have used a running
time of 5 years with a statistics corresponding to a primary proton
beam of $5 \times 10^{20}$ per year, giving a total of $2.5 \times
10^{21}$, the maximum reachable value reported by the MINOS
collaboration. The mean energy of the neutrino beam is $\langle E
\rangle \simeq 3 \,\mathrm{GeV}$.

The OPERA detector is located at Gran Sasso and gets its beam from
CERN (CNGS). OPERA consists of two parts: a muon tracker and an
emulsion cloud chamber. The latter one is the part which is able to
discern a $\nu_\tau$ charged current interaction by identifying the
subsequent $\tau$-decay. The baseline is $732\,\mathrm{km}$.
Following~\cite{Huber:2004ug,Guler:2000bd,Komatsu:2002sz} we assume a
5 year run with a nominal beam intensity of $4.5 \times
10^{19}\,\mathrm{pot}$ per year. The CNGS neutrino beam has an average
energy of $\langle E \rangle \simeq 17 \,\mathrm{GeV}$ .

Since both MINOS and OPERA have the same baseline we use the same
matter density which we take constant and equal to its value at the
Earth's crust, that is $\rho = 2.7$ g/cm$^3$.

Finally, DoubleCHOOZ is a reactor experiment, to be located in the old
site of CHOOZ, in France. The experiment consists of a pair of nearly
identical near and far detectors, each with a fiducial mass of
$10.16\,\mathrm{t}$ of liquid scintillator. The detectors are located
at a distance of $0.2\,\mathrm{km}$ and $1.05\,\mathrm{km}$
respectively. As considered in~\cite{Huber:2006vr} we assume the
thermal power of both reactor cores to be 4.2 GW and a running time of
5 years. The neutrinos mean energy is $\langle E \rangle \simeq 4
\,\mathrm{MeV}$.

Concerning the neutrino oscillation parameters used to calculate the
simulated event rates, we have taken the current best fit values given
in Ref.~\cite{Maltoni:2004ei}, unless stated otherwise:
\begin{equation}
  \begin{array}{rclrclrcl}
    \sin^2\theta^{\rm true}_{12} & = & 0.32, & \sin^2\theta^{\rm true}_{23} & = & 0.5, & \sin^2\theta^{\rm true}_{13} & = & 0,\\
    (\Delta m^2_{21})^{\rm true} & = & +7.6 \times 10^{-5} ~\mathrm{eV^2}, & (\Delta m^2_{31})^{\rm true} & = & +2.4 \times 10^{-3} ~\mathrm{eV^2}, &\delta^{\rm true}_{CP} & = & 0\,.
  \end{array}
\end{equation}
Note the positive sign assumed for $(\Delta m^2_{31})^{\rm true}$
which corresponds to the case of normal hierarchy. Since, none of the
experiments considered here is very sensitive to ordinary matter
effects, our results would be very similar when choosing as true
hierarchy, the inverted one.

\section{Results} 

\subsection{ Disappearance - Probing NU NSI ($\mathbf{\varepsilon_{\tau\tau}}$) }

\begin{figure}[t!]
\begin{center}
\includegraphics[width=0.5\textwidth]{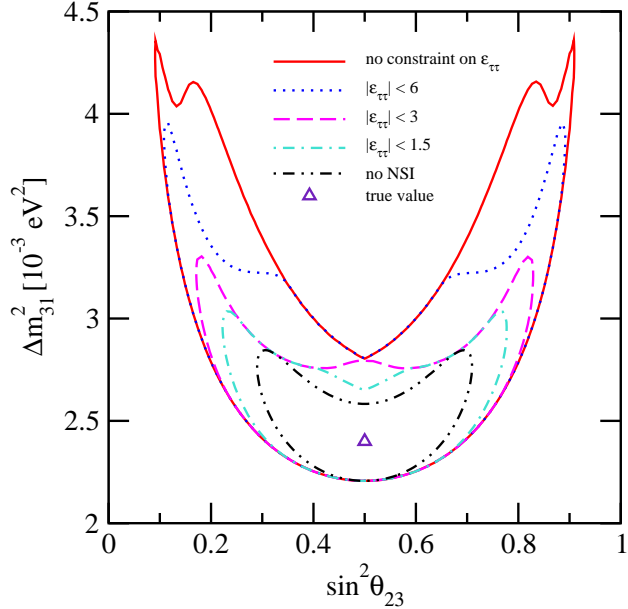}
\caption{\label{fig:atm} Shown is the allowed region in the
  $\sin^22\theta_{23}$-$\Delta m^2_{31}$-plane at 95\% CL (2 dof). In
  this fit $\theta_{13}$, $\varepsilon_{e\tau}$ and $\varepsilon_{\tau\tau}$
  are left free. The different lines correspond to different values
  for $\varepsilon_{\tau\tau}$ as explained in the legend.}
\end{center}
\end{figure}

As it has been previously shown in~\cite{Friedland:2006pi,
  Blennow:2007pu} the presence of NSI, notably $\varepsilon_{\tau\tau}$,
substantially degrades the goodness of the determination of the
``atmospheric'' neutrino oscillation parameters from
experiment. Indeed as shown in figure~\ref{fig:atm} our calculation
confirms the same effect, showing how the allowed region in the
$\sin^2\theta_{23}$-$\Delta m^2_{31}$-plane increases in the presence
of NSI.

This figure is the result of a combined fit to simulated OPERA and
MINOS data in terms of the ``atmospheric'' neutrino oscillation
parameters, leaving the mixing angle $\theta_{13}$ to vary freely. The
inner black dot-dashed curve corresponds to the result obtained in the
pure oscillation case (no NSI).  As displayed in the figure, allowing
for a free nonzero strength for NSI parameters $\varepsilon_{\tau\tau}$
and $\varepsilon_{e\tau}$ the allowed region grows substantially, as seen
in the solid, red curve.
Intermediate results assuming different upper bounds on
$|\varepsilon_{\tau\tau}|$ strengths are also indicated in the figure,
and given in the legend.
One sees that the NSI effect is dramatic for large NSI magnitudes.
However, such large values are in conflict with atmospheric neutrino
data~\cite{Fornengo:2001pm,Friedland:2006pi}.
In contrast, for lower NSI strengths allowed by the atmospheric +
MINOS data combination~\cite{Friedland:2006pi}, say
$|\varepsilon_{\tau\tau}|=1.5$, the NSI effect becomes much smaller.
Clearly beam experiments currently can not compete with atmospheric
neutrino data in constraining $\varepsilon_{\tau\tau}$. The reason for
the good sensitivity of atmospheric data to the presence of NSI is the
very large range in $L/E$, especially the very high energy events are
crucial in constraining NSI~\cite{Fornengo:2001pm}.

In summary, the inclusion of OPERA data helps only for very large
values of $\varepsilon_{\tau\tau}$ as can be seen also from the first
line of table~\ref{tab:00}. These large values, however are already
excluded by the combination of MINOS and atmospheric
results~\cite{Friedland:2006pi}.  We checked that doubling the OPERA
exposure does not change this conclusion. The slight improvement by
OPERA is exclusively due the $\nu_\mu$ sample in the muon tracker and
the results do not change if we exclude the $\nu_\tau$ sample from the
analysis. The usefulness of the $\nu_\mu$ sample stems from the very
different value of $L/E$ compared to MINOS. These results are not too
surprising, since even a very high energy neutrino factory will not be
able to improve the bound on $\varepsilon_{\tau\tau}$ in comparison to
atmospheric neutrino data~\cite{Huber:2001zw}.

\begin{table}[b!]
  \begin{tabular}{|c|c|c|c|c|c|c|}
   
    \hline
    \hline
    & \multicolumn{2}{|c|}{M2} & \multicolumn{2}{|c|}{O} & \multicolumn{2}{|c|}{M2+O} \\
    \hline
    & 90\% C.L. & 95\% C.L. & 90\% C.L. & 95\% C.L. & 90\% C.L. & 95\% C.L. \\
    \hline
    \hline
    $\varepsilon_{\tau\tau}$ & [-10.8,10.8] & [-11.8,11.8] & [-10.4,10.4] & [-11.0,11.0] & [-8.5,8.5] & [-9.2,9.2] \\
    \hline
    $\varepsilon_{e\tau}$ & [-1.9,0.9] & [-2.3,1.0] & [-2.1,1.4] & [-2.5,1.6] & [-1.6,0.9] & [-2.0,1.0] \\
    \hline
    $\Delta m^2_{31}$ [10$^{-3}$ eV$^2$] & [2.3,4.5] & [2.2,4.9] & [2.0,5.0] & [2.0,5.3] & [2.3,3.8] & [2.2,4.0] \\
    \hline
    $\sin^2\theta_{23}$ & [0.08,0.92] & [0.07,0.93] & [0.08,0.92] & [0.07,0.93] & [0.12,0.88] & [0.11,0.89] \\
    \hline
    \hline
  \end{tabular}
  \caption{\label{tab:00}%
    90\% and 95\% C.L. allowed regions for
    $\varepsilon_{\tau\tau}$, $\varepsilon_{e\tau}$, $\Delta m^2_{31}$
    and $\sin^2\theta_{23}$ for different sets of experiments. Each row
    is obtained marginalizing over the remaining parameters in the table, 
   plus $\theta_{13}$. The true value for $\sin^22\theta_{13}$ is  $0$. }
\end{table}

\subsection{Appearance - probing FC NSI ($\mathbf{\varepsilon_{e\tau}}$)}

It is well known that, in the presence of NSI, the determination of
$\theta_{13}$ exhibits a continuous degeneracy~\cite{Huber:2001de}
between $\theta_{13}$ and $\varepsilon_{e\tau}$ which leads to a drastic
loss in sensitivity in $\theta_{13}$.  A measurement of only
$P_{e\mu}$ and $P_{\bar\mu \bar e}$ at one $L/E$ cannot disentangle
the two and will only yield a constraint on a combination of
$\theta_{13}$ and $\varepsilon_{e\tau}$.  In this context, it has been
shown in~\cite{Campanelli:2002cc}, that even a very rudimentary
ability to measure $P_{\mu\tau}$ may be sufficient to break this
degeneracy. Therefore, it seems natural to ask whether OPERA can
improve upon the sensitivity for $\varepsilon_{e\tau}$ that can be
reached only with MINOS. The latter has been studied
in~\cite{Friedland:2006pi} in combination with atmospheric neutrinos
and on its own in Ref.~\cite{Blennow:2007pu}. The result, basically,
was that MINOS will not be able to break the degeneracy between
$\theta_{13}$ and $\varepsilon_{e\tau}$ and hence a possible
$\theta_{13}$ bound from MINOS will, in reality, be a bound on a
combination of $\varepsilon_{e\tau}$ and $\theta_{13}$.

In table~\ref{tab:00} we display our results for a true value of
$\theta_{13}=0$ and no NSI. The allowed range for $\varepsilon_{e\tau}$
shrinks only very little by the inclusion of OPERA data. As in the
case of $\varepsilon_{\tau\tau}$ we explicitly checked that this result
is not due to the $\nu_\tau$ sample in OPERA but is entirely due to
the different $L/E$ compared to MINOS. Also a two-fold increase of the
OPERA exposure does not substantially alter the result.

\begin{figure}
\begin{center}
\includegraphics[width=0.5\textwidth]{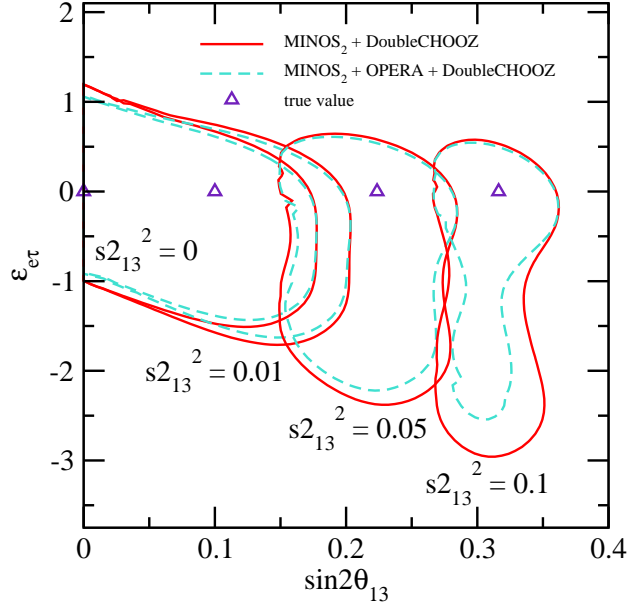}
\caption{\label{fig:th13} Shown are the allowed regions in the
  $\sin2\theta_{23}$-$\varepsilon_{e\tau}$-plane at 95\% CL (2 dof).
  $\Delta m^2_{31}$, $\theta_{23}$ and $\varepsilon_{\tau\tau}$ are left
  free in this fit. The solid lines correspond to the combination of
  MINOS$_2$ and DoubleCHOOZ while the dashed lines also include OPERA
  in the analysis. Each set of lines correspond to different true
  values for $\sin^22\theta_{13}$, from left to right: 0, 0.01, 0.05
  and 0.1.}
\end{center}
\end{figure}

In order to improve the sensitivity to NSI and to break the degeneracy
between $\theta_{13}$ and $\varepsilon_{e\tau}$ it will be necessary to
get independent information on either $\varepsilon_{e\tau}$ or
$\theta_{13}$. An improvement of direct bounds on $\varepsilon_{e\tau}$
is in principle possible by using a very high energy $\nu_e$ beam and
a close detector, but this would require either a neutrino factory or
a high $\gamma$ beta beam. Both these possibilities are far in the
future and will therefore not be considered any further in this
letter. Thus, we focus on independent information on
$\theta_{13}$. Reactor experiments are very sensitive to $\theta_{13}$
but do not feel any influence from $\varepsilon_{e\tau}$ since the
baseline is very short and the energy very low which leads to
negligible matter effects. This is true for standard MSW-like matter
effects as well as non-standard matter effects due to
NSI~\cite{Valle:1987gv}. We consider here as new reactor experiment
DoubleCHOOZ~\cite{Ardellier:2006mn}, but for our discussion Daya
Bay~\cite{Guo:2007ug} or RENO~\cite{Joo:2007zzb} would work equally
well. In figure~\ref{fig:th13} we show the allowed regions in the
$\sin2\theta_{13}$-$\varepsilon_{e\tau}$ plane for the combinations of
MINOS and DoubleCHOOZ (red solid curves) and of MINOS, DoubleCHOOZ and
OPERA (blue dashed curves) for four different input values of
$\sin^22\theta_{13}$ indicated in the plot. As expected, the effect of
DoubleCHOOZ in all four cases is to constrain the allowed
$\sin2\theta_{13}$ range. The impact of OPERA, given by the difference
between the solid and dashed lines, is absent for very small true
values of $\sin2\theta_{13}$ and increases with increasing true
values. For the largest currently permissible values of
$\theta_{13}\simeq0.16$, OPERA can considerably reduce the size of the
allowed region and help to resolve the degeneracy. In that parameter
region a moderate increase in the OPERA exposure would make it
possible to constrain large negative values of
$\varepsilon_{e\tau}$. Again, this effect has nothing to do with
$\nu_\tau$ detection and, in this case, is based on the different
$L/E$ in $\nu_e$-appearance channel.

\begin{table}
  \begin{tabular}{|c|c|c|c|c|c|c|c|c|}
    \hline
    \hline
    & \multicolumn{2}{|c|}{M2} & \multicolumn{2}{|c|}{O} & \multicolumn{2}{|c|}{M2+O} & \multicolumn{2}{|c|}{M2+O+DC} \\
    \hline
    & 90\% C.L. & 95\% C.L. & 90\% C.L. & 95\% C.L. & 90\% C.L. & 95\% C.L. & 90\% C.L. & 95\% C.L. \\
    \hline
    \hline
    $\varepsilon_{\tau\tau}$ & [-10.1,11.0] & [-11.2,12.0] & [-10.1,10.3] & [-10.8,11.0] & [-7.9,9.0] & [-8.7,9.6] & [-5.1,5.3] & [-5.6,5.8] \\
    \hline
    $\varepsilon_{e\tau}$ & [-4.2,1.3] & [-4.5,1.5] & [-4.3,1.5] & [-5.0,1.8] & [-3.7,1.2] & [-4.1,1.4] & [-0.5,0.4] & [-0.7,0.5] \\
    \hline
    $\Delta m^2_{31}$ [10$^{-3}$ eV$^2$] & [2.3,4.6] & [2.2,5.0] & [2.0,4.8] & [2.0,5.2] & [2.3,4.0] & [2.2,4.2] & [2.3,2.8] & [2.3,2.9] \\
    \hline
    $\sin^2\theta_{23}$ & [0.09,0.92] & [0.08,0.93] & [0.09,0.93] & [0.08,0.94] & [0.13,0.90] & [0.12,0.91] & [0.24,0.78] & [0.22,0.80] \\
    \hline
    \hline
  \end{tabular}
  \caption{\label{tab:01}%
    Same as table~\ref{tab:00} with true value $\sin^22\theta_{13}$ of $0.1$. }
\end{table}

\section{Conclusion}

In this letter we have studied how OPERA can help in improving the
sensitivities on neutrino non-standard contact interactions of the
third family of neutrinos.  In our analysis we considered a combined
OPERA fit together with high statistics MINOS data, in order to obtain
restrictions on neutrino oscillation parameters in the presence of
NSI.  Due to its unique ability of detecting $\nu_\tau$ one would
expect that the inclusion of OPERA data would provide new improved
limits on the universality violating NSI parameter
$\varepsilon_{\tau\tau}$. We found, however, that the $\nu_\tau$ data
sample is too small to be of statistical significance. This holds even
if we double the nominal exposure of OPERA to
$9\times10^{19}\,\mathrm{pot}$.  OPERA also has a $\nu_\mu$ sample,
which can help constraining NSI.  Here the effect is due to the very
different $L/E$ of OPERA compared to MINOS. This makes the OPERA
$\nu_\mu$ sample more sensitive to NSI. However, the improvement is
small and happens in a part of the NSI parameter space which is
essentially excluded by atmospheric neutrino data.

We have also studied the possibility of constraining the FC NSI
parameter $\varepsilon_{e\tau}$. For this purpose it is crucial to have a
good knowledge of $\theta_{13}$.  Therefore, we included future
DoubleCHOOZ data, since reactor neutrino experiments are insensitive
to the presence of NSI of the type considered here. Therefore, reactor
experiments can provide a clean measurement of $\theta_{13}$, which in
turn can be used in the analysis of long baseline data to probe the
NSI. DoubleCHOOZ is only the first new reactor experiment and more
accurate ones like Daya Bay or Reno will follow. Our result would be
qualitatively the same if we would have considered those, more
precise, experiments, but clearly the numerical values of the obtained
bounds would improve. The conclusion for $\varepsilon_{e\tau}$ with
respect to the $\nu_\tau$ sample is the same as before: the sample is
very much too small to be of any statistical significance. OPERA's
different $L/E$ again proves to be its most important feature and
allows to shrink the allowed region on the
$\sin^2\theta_{13}$-$\varepsilon_{e\tau}$ plane for large $\theta_{13}$
values. Here a modest increase in OPERA exposure would allow to
completely lift the $\theta_{13}$-$\varepsilon_{e\tau}$ degeneracy and
thus to obtain a unique solution.

\acknowledgments

We would like to thank C. Hagner for useful information about the
OPERA experiment. PH acknowledges the warm hospitality at IFIC at
which parts of this work were performed.  This work has been supported
by Spanish grants FPA2005-01269 (MEC) and ACOMP07/270 (Generalitat
Valenciana) and by the European Commission RTN Contract
MRTN-CT-2004-503369.  AEP thanks CERN Theory division for hospitality
during his stay and MEC for a FPU grant.

% \bibliographystyle{apsrev}
% \bibliography{./references}

 % \bibliographystyle{h-physrev4} 
%  \bibliography{cosm-ref-lattanzi,valle-ref,references-huber}

\end{document}